\documentstyle[12pt,epsfig,psfig]{article}
\newcommand{\be}{\begin{equation}}
\newcommand{\ee}{\end{equation}}
\newcommand{\bea}{\begin{eqnarray}}
\newcommand{\eea}{\end{eqnarray}}

\begin{document}
\vspace*{2cm}
\begin{center}
{\Large\bf Observing compact quark matter droplets in relativistic
nuclear collisions}
\\[2cm]
{\bf K.\ Paech$^a$, J.\ Brachmann$^a$, M.\ A.\ Lisa$^b$, A.\ Dumitru$^c$, 
H.\ St\"ocker$^a$, W.\ Greiner$^a$}
\\
{\small $^a$ Institut f\"ur Theoretische Physik, Universit\"at Frankfurt a.M.,
Postfach 111932,\\D-60054 Frankfurt am Main, Germany\\ }
{\small $^b$ Department of Physics, The Ohio State University, 174 W.\ 18th 
Avenue,\\ Columbus, Ohio 43210, USA}\\
{\small $^c$ Department of Physics, Columbia University, \\
538 W.\ 120th Street, New York, NY 10027, USA}
\\[1cm]
{\large Oct.\ 2000}
\end{center}
\vspace*{1.5cm}
\begin{abstract}
Compactness is introduced as a new method to search for the onset
of the quark matter
transition in relativistic heavy ion collisions.
That transition supposedly leads to stronger compression and higher
compactness of the source in coordinate space. That effect could be
observed via pion interferometry.
We propose to measure the compactness of the source in the
appropriate principal axis frame of the compactness tensor
in coordinate space.
\end{abstract}
\newpage

A variety of
signatures for the observation of the quark matter transition in relativistic
heavy ion collisions have been proposed, see e.g.~\cite{QM99}
and references therein.
Nowadays it is widely believed that
irregularities due to such a new state of matter should be seen most
clearly in excitation
functions, i.e.\ in the bombarding energy dependence, or possibly
also in the impact parameter or mass-number dependence of various
observables.

However, in spite of circumstantial evidence for such a
novel state of matter~\cite{Heinz:2000bk}, 
undisputable ``irregularities'' in such
experimental excitation functions have not been reported to date. Most probably
a convincing argument
that quark matter has indeed been found requires
several independent measurements of distinct observables pointing to the same
conclusion - namely the onset of deconfinement at a particular
bombarding energy, impact parameter and system size.
 
In the present paper we introduce a new observable for the onset of the phase
transition.
It relies on the measurement of the compactness of the source, which is
related to the pressure and density
of the system in the compression and expansion stage of
the nucleus-nucleus collision. Observables related to the pressure
do not only reflect the transient high density but rather relate to the
{\em order parameter} (field), specifying in which thermodynamical 
{\em state}~\cite{LeeWick} the excited matter was created. 
The compactness can be identified via interferometry:
The illumination of the baryon source by the pion radiation is subject to
experimental scrutinity via pion interferometry measurements~\cite{LISA}.

To illustrate the basic idea, let us first discuss compression shocks
in heavy-ion collisions employing the Rankine Hugoniot Taub shock
adiabat (RHTA) solution~\cite{Baumgardt:1975qv,Rischke:1995mt}
of relativistic hydrodynamics,
\be \label{RHTA}
W^2 - {W_0}^2 + P\left(\frac{W}{\rho} - \frac{W_0}{\rho_0}\right) = 0\quad.
\ee
It relates the energy per baryon in the local rest frame of the compressed
matter, $W$, to its baryon density $\rho$ and
pressure $P$. $W_0$ and $\rho_0$ are the energy (=923~MeV) and
density (=0.16~fm$^{-3}$) in the ground state, and the compression factor is
$\rho/\rho_0$.

Solving for $P$ we obtain
\be 
P= W \rho_0 \frac{{\rm \gamma_{cm}}-1/{\rm \gamma_{cm}}}
            {1 - {\rm \gamma_{cm}} \rho_0/\rho} \quad.
\ee
Simply speaking, the onset of the transition to quark matter at a given
incident energy $E^{\rm {kin}}_{\rm {lab}} = 2({\rm \gamma_{cm}}^2 - 1) W_0$
lets the pressure increase less rapidly with $W$ and/or $\rho$, and
consequently higher compression $\rho/\rho_0$ can be achieved than for the
case without transition. Now, as $\rho V\simeq\pi R_{\rm A}^2 L\rho$
must equal $2A$ by virtue of baryon number conservation, the longitudinal
thickness $L$
of the compressed matter is proportional to $1/\rho$. Thus, a transition to
quark matter leads to a more compact system, just as compact stars with
a quark matter core are more compact than pure neutron
stars~\cite{Glendenning:1997wn}. Of course, in heavy-ion collisions that
expectation is based on the behavior of relativistic compression shocks rather
than hydrostatic and gravitational equilibrium.

To investigate quantitatively the experimental observable, we perform
3-dimensional (1-fluid) calculations of relativistic hydrodynamics for the
compression as well as for the subsequent expansion stage of the reaction. That
is,
we solve numerically the continuity equations for the energy-momentum tensor,
$\partial_\mu T^{\mu\nu}=0$, and the net baryon current,
$\partial_\mu N_B^{\mu}=0$.
Detailed discussions of (3+1)-d numerical solutions for
hydrodynamical compression and expansion can be found e.g.\
in~\cite{numH}. We shall employ two different equations of
state (EoS) $P(W,\rho)$:
i) a relativistic mean field (RMF)
hadron fluid~\cite{Serot:1986ey} corresponding
to baryons and antibaryons interacting via exchange of massive scalar and
vector bosons, plus free thermal pions; the parameters of the lagrangian are
fitted to the ground state of infinite nuclear matter, in particular the
nuclear saturation density, the energy per nucleon, and the incompressibility.
ii) the same EoS as in i) for the low-density phase, supplemented by
the Bag Model equation of state with bag constant $B^{1/4}=235$~MeV for the
quark-gluon (QG) phase. The phase coexistence region corresponding to this
first-order transition is constructed employing Gibbs' condition of
phase equilibrium, $p_{\rm RMF}(T,\mu_{\rm B})=p_{\rm QG}(T,\mu_{\rm B})$, where
$T$ and $\mu_B$ denote the temperature and the baryon-chemical potential,
respectively. For example, for $\rho=0$ we find $T_C\approx170$~MeV, while
at $T=0$ phase coexistence sets in at $\rho\approx4.6\rho_0$.
A more detailed discussion of these EoS can be found in~\cite{Rischke:1995mt}.
 
As an example, we study the compactness in the reactions Au+Au at 
impact parameter $b = 3$~fm and energy $E^{\rm kin}_{\rm lab}=
8A$~GeV, where  the transition to quark matter does
occur (within the present model).
We define the configuration space sphericity tensor, also called
``compactness tensor'' below, as the second moment of the net baryon current.
On fixed time hypersurfaces we have
\be
F_{ij} (t) = \int {\rm d}^3x \,\,  x_i\, x_j \,\, N_B^0(t,\vec{x}) \,\,
\Theta\left(\rho(t,\vec{x})- \rho_{cut}\right) \quad.
\ee
We apply an additional density cut $\rho>\rho_{cut}$ in the integral to discard
spectator matter. In the future the cuts and the hypersurface will have to be
adapted to the experimental conditions. However, that is not crucial for
understanding the effect.

The three eigenvalues $f_n$ are the solutions of the cubic equation
$det (F_{ij}-f\delta_{ij}) = 0$, and the eigenvectors 
$\vec{e}_n$ follow from solving the
linear system of equations $(F_{ij}-f_n\delta_{ij})e_n^j = 0$.
Let $\vec{e}_1$, $\vec{e}_2$ be the eigenvectors defining the reaction
plane, $\vec{e}_1$ corresponding to the bigger of the eigenvalues
$f_1$ and $f_2$. To simplify the discussion we shall assume that
the matter distribution is
symmetric with respect to the reaction plane.

We can now rotate the coordinate frame around $\vec{e}_3$ by an angle
$\cos\Theta=\vec{e}_1\cdot\vec{e}_z$, where $\vec{e}_z$ defines the 
longitudinal (beam-) direction in the lab frame.
The rotated compactness tensor $F^*$ can be written
in terms of the eigenvalues $f_n$ and orthogonal eigenvectors
$\vec{e}_n^{\, *}$ as
\be \label{diagF}
F^* = f_1 \vec{e}_1^{\, *} \otimes\vec{e}_1^{\, *} 
  + f_2 \vec{e}_2^{\, *} \otimes\vec{e}_2^{\, *}
  + f_3 \vec{e}_3^{\, *} \otimes\vec{e}_3^{\, *}\quad.
\ee
In diagonal form, $F^*$ specifies an ellipsoid in configuration space with
principal axis along $\vec{e}_n^{\, *}$ and radii $\sqrt{f_n}$.
For example, cigar patterns, oriented along the z-axis, would correspond
to $f_1>f_2=f_3$,
$\vec{e}_1^{\, *}=\vec{e}_z$, $\vec{e}_2^{\, *}=\vec{e}_x$, $\vec{e}_3^{\, *}
=\vec{e}_y$.
The ``compactness'' can now be defined as the ratio of the in-plane
radii, $\sqrt{f_2/f_1}$.

\begin{figure}[htp]
\hspace{0cm}\centerline{\hbox
{\epsfig{figure=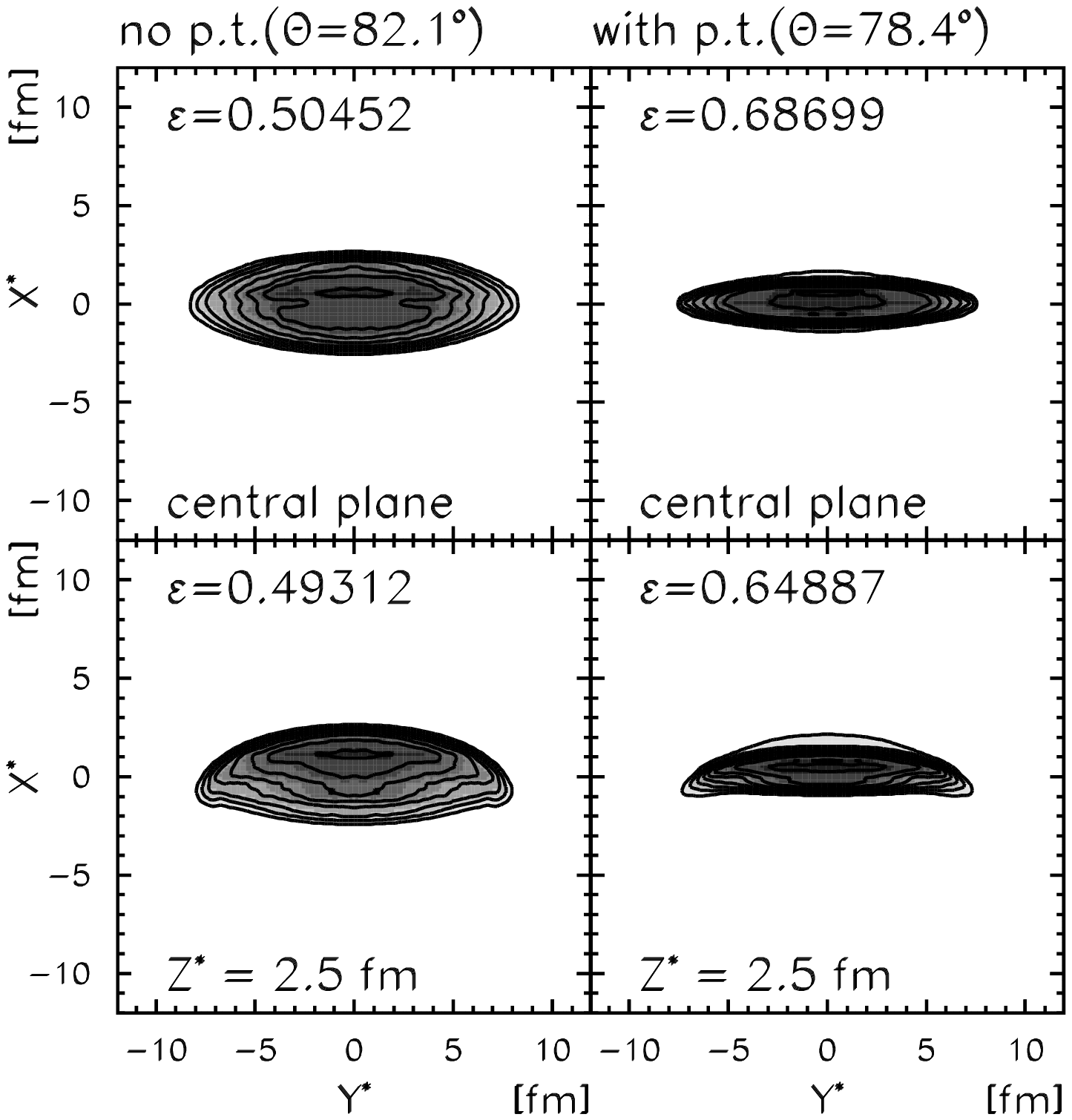,width=10cm}}}
\caption{Contour plot of the time-like component of the net baryon
four-current in planes transverse to the longitudinal axis $\vec{e}_1^{\, *}$
 of the
compactness tensor (Au+Au collision at $E^{kin}_{lab}=8$~AGeV, $b=3$~fm).
Left: Relativistic mean-field EoS without phase transition. Right:
with phase transition to quark matter.}
\label{fig1}
\end{figure}
\begin{figure}[h]
\hspace{0cm}\centerline{\hbox
{\epsfig{figure=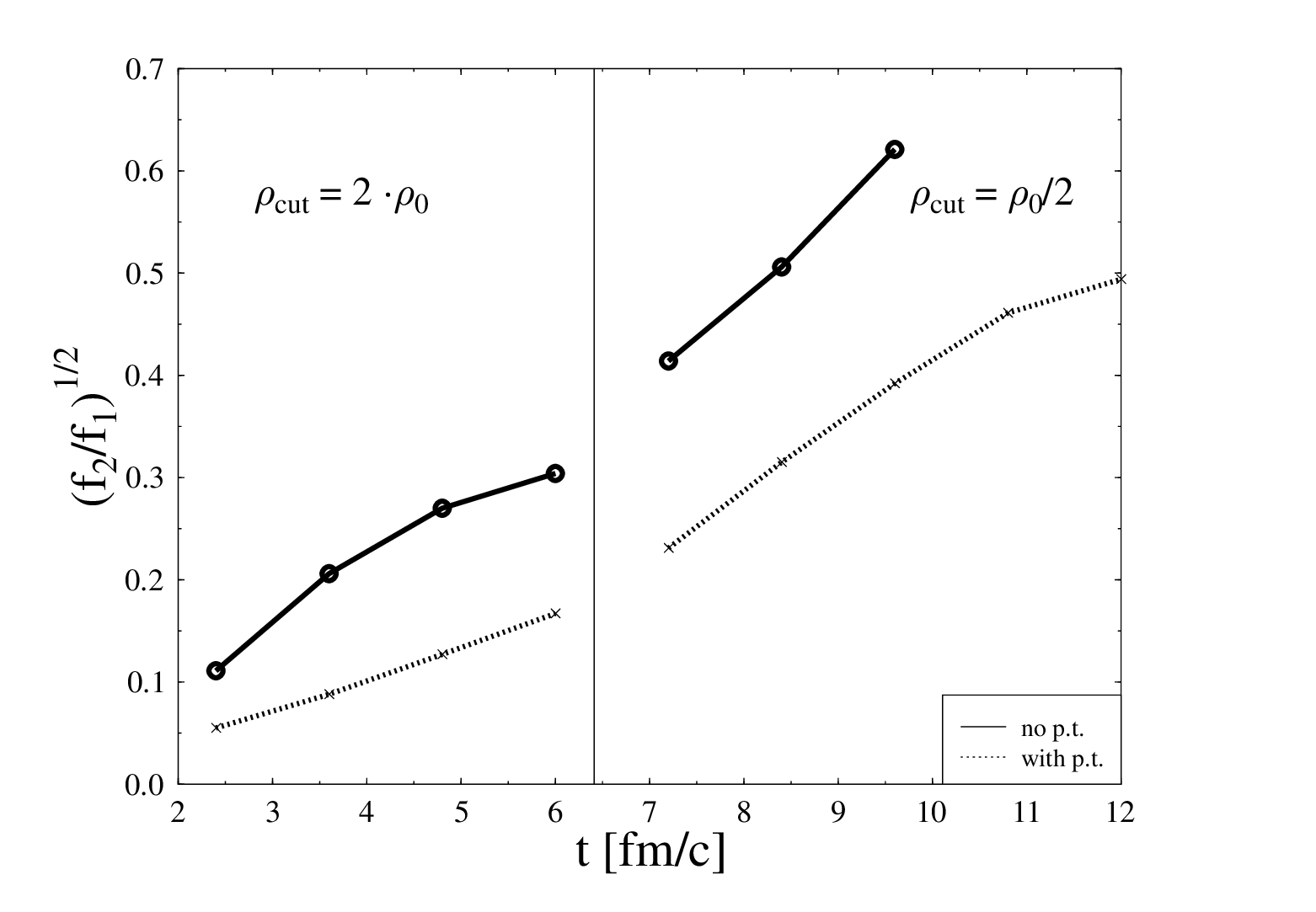,width=12cm}}}
\caption{Ratio of the in-plane radii $\sqrt{f_2/f_1}$ as it evolves
during the compression and expansion stages, for the relativistic
mean-field EoS without phase transition and for the case with transition
to quark matter (Au+Au, $E^{kin}_{lab}=8$~AGeV, $b=3$~fm).}
\label{fig2}
\end{figure}

Fig.~\ref{fig1} shows typical density distributions in the transverse planes
of the coordinate frame where $F^*$ is diagonal, eq.~(\ref{diagF}).
The polar angle relative to the lab frame as well as the eccentricity
\be
\epsilon = \frac{f_3-f_2}{f_3+f_2}
\ee
are shown as well. 
As already indicated in the introduction, we find very different eigenvalues
for the two equations of state, the one without and the other with a
first order phase transition included.
The calculation with transition to quark
matter corresponds to higher compactness of the baryon distribution.
Because the incompressibility $\partial p/\partial\rho$ or
$\partial p/\partial(W\rho)$ is smaller in the presence of the
coexistence phase,
the compactness tensor is much flatter (nearly a factor of two~!)
in model ii) than in model i).
Moreover, our (3+1)-dimensional expansion solutions show that
the ratio of the in-plane
radii $\sqrt{f_2/f_1}$ remains much smaller in the case with phase
transition, cf.\ Fig.~\ref{fig2}.

This allows to measure directly the density increase in the high density stage
of the reaction, if a phase transition occurs. Care must be taken that
the investigated range of impact parameters constitutes a
moderately small bin of centrality values. One should keep in mind
that the compression factor is affected by the incompressibility
evaluated on the RHTA~(\ref{RHTA}), but
{\em not} along a path of fixed specific entropy, because a large amount
of entropy is being produced in the compression process.

The predicted change in compactness due to the reduced incompressibility
associated with a first-order phase transition
can be observed experimentally via
the novel interferometry analysis developed recently by Lisa {\it et
al}.~\cite{LISA}.
The proposed method is quite robust and incorporates other interesting
information as the configuration space tilt angles~\cite{Brachmann:2000xt},
which complement the momentum-space flow angles.
It avoids cuts in tilted ellipsoids, which are not
analysed in the appropriate rotated frame, and were the eccentricity and
the RMS-radii are much less distinct for models i) and ii).

{\bf Acknowledgements:}
This work is supported by BMBF, GSI, DFG, and Graduiertenkolleg ``Theoretische
und Experimentelle Schwerionenphysik''.
A.D.\ acknowledges support from the DOE Research Grant, Contract No.\
De-FG-02-93ER-40764.
We would like to thank Stefan Scherer, Sven Soff and Steffen Bass for
stimulating discussions.

\end{document}